\documentclass[a4paper,11pt]{article}

\usepackage{psfig}

\def\msun {\hbox{M$_{\odot}$}} 
\def\tiff {$^{44}$Ti} 
\def\scff {$^{44}$Sc}
\def\caff {$^{44}$Ca}
\def\degr{\hbox{$^\circ$}}
\def\arcmin{\hbox{$^\prime$}}

\begin{document}

\vspace{1.0cm}
{\Large \bf THE  HARD X-RAY EMISSION AND $^{44}$\bf Ti EMISSION OF CAS A}

\vspace{1.0cm}

Jacco Vink, Jelle S.  Kaastra, Johan A. M. Bleeker, and Hans Bloemen

\vspace{1.0cm}

{\it SRON, Sorbonnelaan 2 NL-3584 CA Utrecht, The Netherlands}\\

\vspace{0.5cm}

\section*{ABSTRACT}
We present an analysis of the BeppoSAX high X-ray energy spectrum
of the supernova remnant Cassiopeia A with an observation time 
of $83\, 10^3$~s. 
We measure a flux upper limit of $4.1 \ 10^{-5}$~ph cm$^{-2}$ s$^{-1}$ 
(99.7\% confidence) of the nuclear decay 
lines of \tiff\ at 68~keV and 78~keV that is lower and inconsistent with 
the flux of an accompanying line at 1157~keV measured by CGRO's Comptel. 
However, if the underlying X-ray continuum is lower, 
because the spectrum is steepening, the actual \tiff\ flux may be higher and 
consistent with the Comptel result, although the measured flux of
($2.9 \pm 1.0)\ 10^{-5}$~ph cm$^{-2}$ s$^{-1}$ under this 
assumption is still lower than the flux measured by Comptel.

\section{INTRODUCTION}
The hard X-ray emission of the supernova remnant Cassiopeia A is interesting
for two reasons. One reason is the hard X-ray continuum emission, 
which may be either synchrotron radiation from electrons with energies in 
excess of 10~TeV or non-thermal bremsstrahlung.

\begin{figure}[tb]
\hbox{
		\psfig{figure=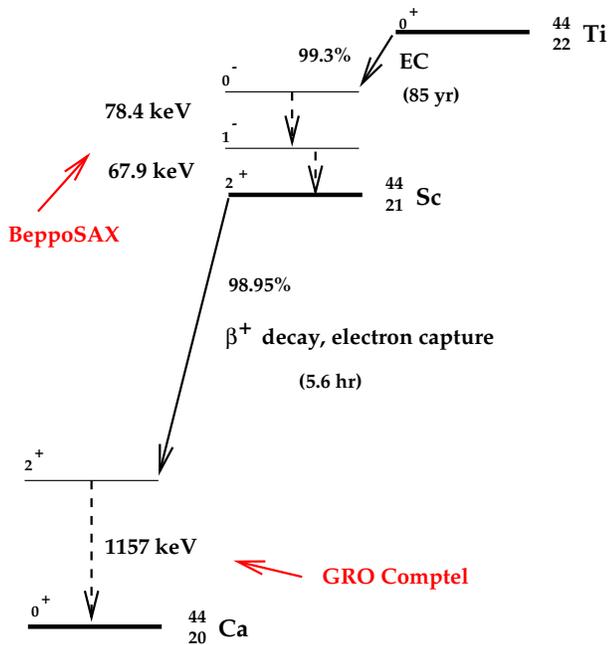,width=8.0cm}
	
	\hfill
	\parbox[b]{75mm}{
	        \caption[]{The decay scheme of \tiff\ with the most important branches.
\label{decay}}
}}
\end{figure}

Equally interesting, but from a totally different perspective, is the
nuclear decay line emission of \tiff\ (for a review see Diehl \& Timmes 1998). 
This unstable element is the product of (explosive) 
nucleosynthesis in core collapse supernovae. 
The production of \tiff\ is closely linked to the production of
$^{56}$Ni, another unstable element. \tiff\ has, however, a longer decay time,
which has only recently been accurately determined to be $85.4\pm 0.9$~yr 
(Ahmad et al. 1998,  Norman et al. 1998, G\"orres et al. 1998).
The decay of \tiff\ to \scff, which decays within a few hours to \caff,
is nearly always accompanied by the emission of 3 photons 
(see figure \ref{decay}) with
energies of 67.9~keV, 78.4~keV and 1157~keV.
The production of \tiff\ and $^{56}$Ni depends on the mass cut in 
the supernova, defining which part of the star will be ejected and
which part of the stellar core will finally end up in the stellar remnant
(neutron star or black hole).

The lifetime of \tiff\ makes that the line emission accompanying its decay
can only be observed when a supernova remnant is young. 
The best candidate in our galaxy is Cassiopeia A (Cas A), 
which is about 320 yr old and is at a distance of  $3.4^{+0.3}_{-0.1}$~kpc 
(Reed et al. 1995). 
The supernova itself must have been an underluminous or an obscured event, 
since no reports of its explosion exist.
However, Ashworth (1980) suggested that an unknown 
6 magnitude star near the position of Cas A
observed by Flamsteed in 1680 may in fact have been the supernova.
Indeed, the  1157~keV is observed by the Comptel experiment on board CGRO 
(Iyudin et al. 1994). 
The most recent flux estimate is 
$(4.8\pm 0.9)\ 10^{-5}$~ph cm$^{-2}$ s$^{-1}$  
(Iyudin et al. 1997), which translates into an initial \tiff\ mass
of $M(^{44}{\rm Ti}) = 2.6\ 10^{-4}$\msun.
Such a large production of \tiff\ is at odds with the lack of X-ray 
iron emission associated with the shocked ejecta (Vink et al. 1996)
and with idea that the supernova may have produced a black hole.
This idea is based upon the apparent lack of a point source, 
pulsar or pulsar nebula in Cas A.
As for the apparent lack of iron (the end product of $^{56}$Ni),
Nagataki et al. (1998), argue that a non-spherically symmetric explosion
may have given rise to a large \tiff\ to $^{56}$Ni ratio.
More accurate observations of \tiff\ and iron may shed new light on 
the nature of the explosion and the mass of the stellar remnant in Cas A.

So far no firm detection of the 67.9~keV and 78.4~keV has been reported,
although some attempts have been made by CGRO (OSSE) (The et al. 1996) 
and RXTE (Rothschild et al. 1997). 
As we shall see, the problem with the detection of these lines is partly the 
uncertain contribution of the continuum radiation in the energy range where
the line emission is expected.

The hard X-ray continuum was first observed by Pravdo \& Smith (1979),
but has only recently been firmly established 
(The et al. 1996, Favata et al. 1997 and 
Allen et al. 1997). 
The current ideas are that it is either synchrotron radiation from shock 
accelerated electrons with energies in the TeV range, in analogy with the 
featureless spectrum of SN1006 (Koyama et al. 1995, 
Reynolds 1998), 
or that it is non-thermal bremsstrahlung  (Asvarov et al. 1990), 
in which case we are dealing with supra-thermal electrons.
Such non-Maxwellian electron distributions are not unlikely given the fact that
electrons with energies in excess of 10~keV have a long relaxation time
(of the order of the age of Cas A) assuming Coulomb interactions 
(see Laming 1998 and Vink et al. 1998). 
Non-thermal bremsstrahlung should exist at some level, since the
cosmic ray electron population has to join the thermal electron distribution,
but at the moment the most likely explanation for the hard X-ray tail is 
synchrotron radiation (Allen et al. 1997, Vink et al. 1998).
An inverse Compton scattering origin of the hard X-ray tail is unlikely,
as shown by Allen et al. (1997), but it should be noted that
Allen et al. did only take into account scattering of 
cosmic microwave background photons, 
whereas Cas~A produces itself a lot of infrared photons
which potentially constitute the dominant source for inverse Compton 
scattering.
The best argument against an inverse Compton origin is
that the spectral photon index of the hard X-ray spectrum should 
be $\Gamma = \alpha + 1$, with $\alpha= 0.79$ the radio spectral index.
This is less steep than $\Gamma = 3.0$, observed by CGRO, RXTE and BeppoSAX.
Inverse Compton emission may, however, be important at higher photon energies.
The location of the hard X-ray emission has recently been 
determined to originate from the Western region of the remnant
(Vink et al. 1998).

\section{INSTRUMENT AND OBSERVATIONS}
Our analysis presented here is mainly based on the Phoswich Detector 
System (PDS, see Frontera et al. 1997a) on board the 
BeppoSAX satellite (Boella et al. 1997), 
which is the most sensitive instrument on board BeppoSAX above 20~keV.
The instrument consists of four detector units and two collimators with a
field of view of 1.3\degr (FWHM). 
Each collimator is rocked between the source and a 210\arcmin\ off 
source position every 94~s (by default). 
The energy resolution above 60~keV is better than 15\%.
The low inclination of the BeppoSAX orbit (3.9\degr) results in a low and stable cosmic ray background and consequently in a relatively low internal 
background.

BeppoSAX observed Cas A five times during the performance verification phase
in August and September 1996. 
An additional observation was made on November 26, 1997. 
The net observation time for the PDS amounted to 82.9~ks.
The reduction of the data was done using the SAXDAS 1.3 reduction package.

A comparison of the hard X-ray flux density of Cas A with HEXTE on board RXTE 
(Allen et al. 1997) and OSSE on board CGRO (The et al. 1996)
and the PDS reveals that there is a flux discrepancy between HEXTE and OSSE on 
the one hand and the PDS on the other hand. 
The flux density observed by the PDS is 40\% lower than OSSE and HEXTE.
Part of the problem is known, since observations of the Crab nebula show that a
correction of 5\% to 15\% should be applied to match the PDS and the other 
narrow field instruments (NFI) on BeppoSAX (Dal Fiume, private communication). 
However, a 15\% correction is not nearly enough to explain the difference 
with OSSE and HEXTE. 
HEXTE and OSSE are detectors similar to the PDS facing the same type of 
calibration errors as the PDS, but the PDS has a lower and more stable
background than HEXTE and OSSE.
Calibration observations of the Crab nebula by HEXTE and the PDS 
do agree in flux within 5\% (Rothschild et al. 1998 and 
Frontera et al. 1997b).
In the analysis presented here we only take into account a 15\% flux 
correction, which is the typical correction needed to match the 
BeppoSAX Medium Energy Concentrator (MECS) results and the PDS.

\begin{table}
	\caption{The best fit parameters. 
Upper limits are 3$\sigma$ (99.7\%, $\Delta\chi^2 = 9.0$) confidence levels.
Note that the thermal component uses abundances similar to 
Favata et al. (1997). 
The temperature of the thermal component was $T_{\rm e} $=4.2~keV\label{table}}
	\begin{flushleft}
{\small
	\begin{tabular}{llll}
        	\hline\noalign{\smallskip}
parameter  & All NFI & PDS & PDS\\
           &         &     & without continuum\\
        	\hline\noalign{\smallskip}
Power law norm (keV$^{-1}$ cm$^{-2}$ s$^{-1}$  @ 1~keV)  
                                   & $1.01 \pm 0.04$  & $0.43 \pm 0.14$ & -- \\
Power law slope ($\Gamma$)         & $3.17\pm 0.02$ &  $2.90 \pm 0.11 $ & -- \\
\tiff\ flux ($10^{-5}$ ph cm$^{-2}$ s$^{-1}$ ) & $1.1\pm 1.1$ ($< 4.6$) & $0.3\pm 1.2$ $(< 4.1)$ & $2.9 \pm 1.0$ ($< 6.4$)\\
$n_{\rm e} n_{\rm H}V (10^{58} {\rm cm}^{-3})$, 
($T_{\rm e} $=4.2~keV) & $1.28 \pm 0.07$   & 1.28 (fixed) & --\\
$\chi^2/\nu$  &  2259/164 \\
$\chi^2/\nu$ (PDS only)   & 22.9/15 & 20.9/16 & 0.51/1  \\
        	\hline\noalign{\smallskip}
	\end{tabular}
}
	\end{flushleft}
\end{table}

\begin{figure}[t]
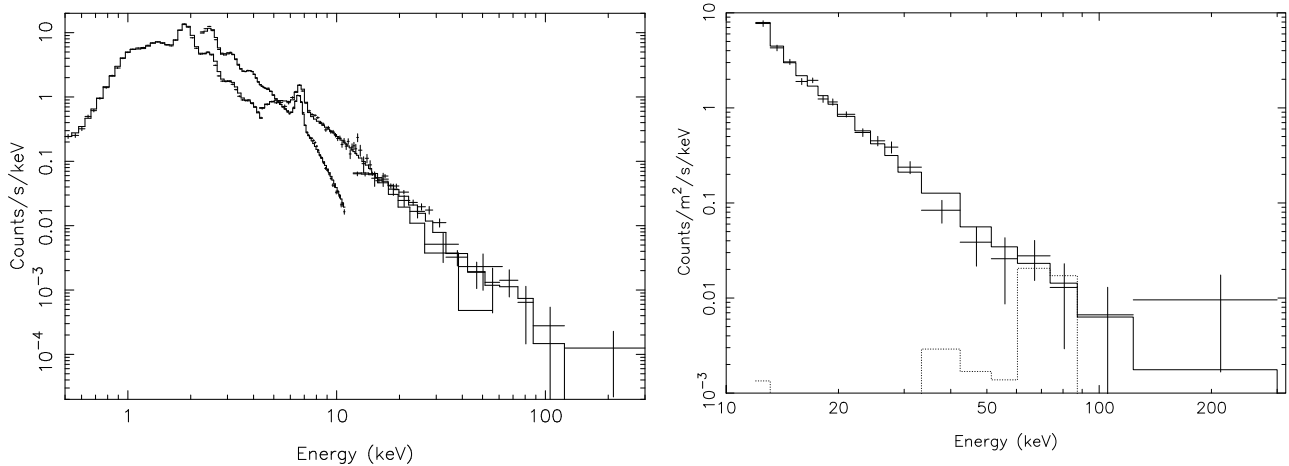

	\centerline{
        	\psfig{figure=vink-fig.2a.ps,width=8.5cm,angle=-90,clip=2}
		\psfig{figure=vink-fig.2b.eps,width=9.3cm,angle=-90,clip=2}
	}
        \caption[]{On the left: The combined BeppoSAX spectrum of Cas A. 
The solid line shows the convolved model.
On the right: The hard X-ray spectrum of Cas A as observed by the PDS. 
The solid line shows the model listed in column 3 of Table~\ref{table}.
The dotted line shows the model listed in column 4  of Table~\ref{table}
(i.e. the model is only fitted to the two bins which have the largest 
response to the two \tiff\ lines.)
\label{nfi}}
\end{figure}
\enlargethispage{\baselineskip}
\section{RESULTS}
We fitted the spectra using the X-ray spectral analysis program SPEX
(Kaastra et al. 1996). We present here the results of
two approaches. 
One is to fit all NFI spectra together with a two temperature plasma 
in non-equilibrium ionization plus a power law and two delta lines at 
67.9~keV and 78.4~keV (cf. Favata et al. 1997); 
an equal flux value for both lines was imposed.
Interstellar absorption corresponding to a hydrogen column density of
$N_{\rm H} = 1.3\ 10^{22}$~cm$^{-2}$ s$^{-1}$ is included 
(Morrison \& McCammon 1983).
The parameters with which we are concerned in this paper
(i.e. the \tiff\ emission and the hard X-ray tail)
are then influenced through various correlations by the observed 
X-ray spectra at low energy, which have a very large statistical weight.
The error on the resulting power law index is small (Table~\ref{table}),
but this does not reflect the uncertainty of the spectral index above 20~keV.

The other approach is to fit only that part of the spectrum in which we are
currently interested, i.e. we only fit the PDS spectrum. 
The problem with this approach is that we need an estimate of the thermal 
contribution below 20~keV. 
The electron temperature of the thermal spectrum is $\sim$ 4~keV. 
This is not only derived from the continuum shape, 
which after all may be influenced by the non-thermal component,
but also from the emission lines of especially He-like and H-like iron.
The fit to the total NFI spectrum does give an estimate of the thermal 
contribution, but it may be unreliable, since we assume that the non-thermal
contribution has a power law shape over the total energy range.
In reality, however, a synchrotron spectrum is likely to bend as the result 
of electron energy losses (Reynolds 1998). Non-thermal 
bremsstrahlung, on the other hand, is probably an extension to the 
thermal spectrum rather than an addition.
In order to see how the thermal contribution affects our results, we
varied the normalization of the thermal component. 
An increase of the normalization by 20\% doubled the best fit \tiff\ flux,
whereas a decrease by 20\% resulted in a negative best fit value for the 
\tiff\ flux.
The value listed in Table~\ref{table} is the value for which the thermal
normalization equals the value obtained for a fit to all NFI spectra,
but since the other narrow field instruments do not constrain the shape of 
the spectrum below 10~keV a wider range of power law index values is allowed.

\begin{figure}[t]
\hbox{
		\psfig{figure=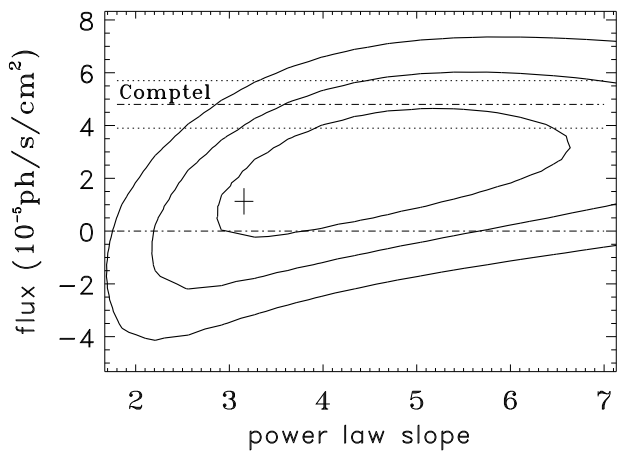,width=8.8cm}
	\hfill
	\parbox[b]{60mm}{
		\caption[]{The error ellipse for the power law slope and \tiff\ flux for a fit to  the PDS data above 30~keV.
The contour lines are for 1, 2 and 3 $\sigma$ errors 
($\Delta \chi^2 = $ 2.3, 6.17, 11.8). The flux measured by Comptel and the 
its 1$\sigma$ errors are also indicated.
\label{ellips2}}
	}}
\end{figure}

Note that the correlation between power law index and \tiff\ flux 
works in two directions: the possible contribution of \tiff\ emission 
results in an uncertain continuum contribution above 50~keV;
it may well be that the hard X-ray continuum
steepens going to higher energies as expected for a synchrotron spectrum 
(Reynolds 1998).
The continuum above 90~keV seen in figure \ref{nfi} is
at the $2\sigma$ level.
The dependence of the \tiff\ flux on the spectral index is shown in 
figure \ref{ellips2}. Only the spectrum above 30~keV has been used for this 
ellipse; the contour levels are for two parameters of interest. 

An absolute upper limit to the \tiff\ is obtained by fitting only that 
part of the spectrum where the \tiff\ line emission is expected and to assume 
that there is no continuum (see figure \ref{nfi}).
The best fit \tiff\ flux under this unlikely assumption
(i.e. ($2.9\pm 1.0)\ 10^{-5}$~cm$^{-2}$ s$^{-1}$) 
is still lower than the best fit value for the 1157~keV line measured by 
Comptel.
This flux translates into an initial \tiff\ mass of 
$M(^{44}{\rm Ti}) = (1.6 \pm 0.7) 10^{-4}$~\msun, the error estimate takes
into account the flux, distance and \tiff\ lifetime uncertainties
and an uncertainty of 20~yr in the age of Cas~A.
The lower \tiff\ flux as measured by the PDS does somewhat alleviate the
problem of the high \tiff\ to $^{56}$Ni ratio.

\section{CONCLUSION}
We have presented an analysis of the X-ray spectrum of Cas A 
from $\sim$ 0.5~keV $\sim 100$~keV as observed with the 
narrow field instruments on board BeppoSAX. 
In this paper we concentrated on the hard X-ray emission observed by 
one of the instruments, the PDS. 
The main interests are the \tiff\ nuclear decay lines, expected at
67.9~keV and 78.4~keV, and the hard X-ray continuum. 
We have shown that the uncertainties in the hard X-ray continuum 
prevent a unique measurement of the \tiff\ flux. 
The uncertainties are mostly in the exact shape of the hard X-ray continuum.
If the spectrum steepens gradually, the \tiff\ may be larger than in the
case of a power law spectrum. 
However, even if we totally discard a continuum contribution we end up 
with a \tiff\ flux measurement of ($2.9 \pm 1.0)\ 10^{-5}$~cm$^{-2}$ s$^{-1}$, 
which is lower than the flux of the 1157~keV \tiff\ line  of ($4.8 \pm 1.0)\ 10^{-5}$~cm$^{-2}$ s$^{-1}$ measured with 
Comptel (Iyudin et al. 1997), 
although the PDS upper limit of $6.4\ 10^{-5}$~cm$^{-2}$ s$^{-1}$
(99.7 \% confidence) is in agreement with the Comptel measurement.
The assumption of no continuum contribution is, however, 
almost certainly wrong and assuming a simple power law spectrum yields
an upper limit to the \tiff\ flux of $4.1\ 10^{-5}$~cm$^{-2}$ s$^{-1}$
(99.7 \% confidence), which is
statistically inconsistent with the value measured by Comptel.

The continuum flux itself is interesting for the information it provides
on cosmic ray physics. If the spectrum indeed steepens above 30~keV this could
be an extra, although inconclusive, argument for a synchrotron origin.
Future observations, for example by Integral, may provide new data on the
emission above 100~keV, which will constrain the actual form of
the hard X-ray spectrum and thus provide us with new clues on the origin of
the spectrum and the \tiff\ flux.

{\small
{\it Acknowledgements}
We gratefully acknowledge helpful discussions with Glenn Allen
and Yuko Mochizuki. 
We thank Fabio Favata for his contribution to the
data reduction and his efforts concerning this project.
This work was financially supported by NWO, 
the Netherlands Organization for Scientific Research.
}

\section{REFERENCES}
\vspace{-5mm}
\begin{itemize}
\setlength{\itemindent}{-8mm}
\setlength{\itemsep}{-1mm}
	\item[]Ahmad, I., Bonino, G., Cini Castagnoli, G., Fischer, S.M., Kutschera, W., Paul, M., Three-Laboratory Measurement of the $^{44}$Ti Half-Life, {\it PhRvL}, {\bf 80}, 2550 (1998).
	\item[]Allen, G.E., Keohane, J.W., Gotthelf, E.V., Petre, R., Jahoda, K., et al., Evidence of X-Ray Synchrotron Emission from Electrons Accelerated to 40 TeV in the Supernova Remnant Cassiopeia A, {\it ApJL}, {\bf 487}, 97 (1997).
	\item[] Ashworth  W.B., A probable Flamsteed observation of the Cassiopeia A supernova, {\it J. Hist. Astr.},{\bf 11}, 1 (1980)
	\item[]Asvarov A.I., Dogiel V.A., Guseinov O.H., Kasumov F.K., The hard X-ray emission of the young supernova remnants, {\it A\&A}, {\bf 229}, 196 (1990).
        \item[]Boella G., Butler, R.C., Perola, G.C., Piro, L., Scarsi, L., Bleeker, J.A.M., BeppoSAX, the wide band mission for X-ray astronomy, {\it A\&AS}, {\bf 122}, 299 (1997).
	\item[]Diehl R., Timmes F.X., Gamma-Ray Line Emission from Radioactive Isotopes in Stars and Galaxies, {\it PASP}, {\bf 110}, 637 (1998).
	\item[]Favata F., Vink J., Dal Fiume D., Parmar, A.N., Santangelo, A., Mineo, T., Preite-Martinez, A., Kaastra, J.S., Bleeker, J.A.M., The broad-band X-ray spectrum of the Cas A supernova remnant as seen by the BeppoSAX observatory, {\it A\&A}, {\bf 324}, L49 (1997).
	\item[]Frontera F., Costa E., Dal Fiume D., Feroci, M., Nicastro, L., et al., The high energy instrument PDS on-board the BeppoSAX X-ray astronomy satellite, {\it A\&AS}, {\bf 122}, 357 (1997a).
	\item[]Frontera F., Costa E., Dal Fiume D., Feroci, M., Nicastro, L., et al., PDS experiment on board the BeppoSAX satellite: design and in-flight performance results, {\it SPIE}, {\bf 3114}, 206, (1997b).
	\item[]G\"orres J., Mei\ss ner Schatz H., Stech, E., Tisschhauser, P., et al., Half-Life of $^{44}$Ti as a Probe for Supernova Models, {\it PhRvL}, {\bf 80}, 2554 (1998).
	\item[]Iyudin A.F., Diehl R., Bloemen H., Hermsen, W., Lichti, G.G., et al., COMPTEL observations of Ti-44 gamma-ray line emission from Cas A, {\it A\&A}, {\bf 284}, L1 (1994).
	\item[]Iyudin  A.F., Diehl R., Lichti G.G.., et al., in proc. of 2nd INTEGRAL workshop, ``The transparant universe'', ed.  (1997).
	\item[]Kaastra J.S, Mewe R., Nieuwenhuijzen H., In: Watanabe, T. (ed.) The 11$^{th}$ coll. on UV and X-ray Spectr. of Astroph. and Laboratory Plasmas, p. 411 (1996).
	\item[]Koyama K., Petre R., Gotthelf E.V., Hwang, U., Matsura, M., Ozaki, M., Holt, S.S., Evidence for shock acceleration of high-energy electrons in the supernova remnant SN1006, {\it Nat}, {\bf 378}, 255 (1995).
	\item[]Laming J.M., A thermal model for the featureless X-ray emission from SN 1006?, {\it ApJ}, {\bf 499}, 309 (1998).
        \item[]Morrison R., McCammon D., {\it ApJ}, {\bf 270}, 119 (1983).
	\item[]Nagataki S., Hashimoto M., Sato K., Yamada, S., Mochizuki, Y.S., The High Ratio of 44Ti/ 56Ni in Cassiopeia A and the Axisymmetric Collapse-driven Supernova Explosion, {\it ApJ}, {\bf 492}, L45 (1998).
	\item[]Norman E.B., Browne E., Chan Y.D., Goldman, I.D.,Larimer, R.-M., et al., Half-life of \tiff, {\it PhRC}, {\bf 57}, 2010 (1998).
	\item[]Pravdo S.H, Smith B.W., X-ray evidence for electron-ion equilibrium and ionization nonequilibrium in young supernova remnants, {\it ApJ}, {\bf 234}, L195 (1979).
        \item[]Reed J.E., Hester J.J., Fabian, A.C, Winkler P.F., The Three-dimensional Structure of the Cassiopeia A Supernova Remnant. I. The Spherical Shell, {\it ApJ}, {\bf 440}, 706 (1995).
	\item[]Reynolds S.P., Models of Synchrotron X-Rays from Shell Supernova Remnants, {\it ApJ}, {\bf 493}, 375 (1998).
	\item[]Rothschild R.E., Lingenfelter R.E., Heindl,W.A., Blanco, P.R., Pelling, M.R, et al., RXTE Observations of Cas A, in Proc. of the 4th Compton Symposium, Dermer, C.D., Strickman, M.S., Kurfess, J.D. (Ed.), AIP Conference Proceedings 410, p. 1089. (1997)
	\item[]Rothschild R.E., Banco P.R., Gruber D.E., Heindl, W.A., MacDonald, D.R., et al., In-Flight Performance of the High-Energy X-Ray Timing Experiment on the Rossi X-Ray Timing Explorer, {\it ApJ}, {\bf 496}, 538 (1998).
	\item[]The L.-S., Leising M.D., Kurfess J.D., Johnson, W.N., Hartmann, D.H., et al., CGRO/OSSE observations of the Cassiopeia A SNR, {\it A\&AS}, {\bf 120}, 357 (1996).
	\item[]Vink J., Kaastra J.S., Bleeker J.A.M., A new mass estimate and puzzling abundances of SNR Cassiopeia A, {\it A\&A}, {\bf 307}, L41 (1996).
	\item[]Vink J., Maccarone, C., Kaastra, J.S., Mineo, T., Bleeker, et al., accepted by A\&A, astro-ph/9902071 (1999).
\end{itemize}

\end{document}